\documentclass[journal]{IEEEtran}

\newcommand{\sqr}{\vrule height6pt width6pt depth1pt}
\def\qed{\hfill\sqr}

\usepackage{graphics} 
\usepackage{epsfig} 
\usepackage{times} 
\usepackage{amsmath} 
\usepackage{amssymb}  
\usepackage{setspace}
\usepackage{color}
\usepackage{fancyhdr}

\definecolor{dred}{rgb}{0.75,0,0}
\definecolor{mred}{rgb}{0.75,0,0}
\definecolor{dgreen}{rgb}{0,0.75,0}
\definecolor{mgreen}{rgb}{0,0.75,0}
\definecolor{dblue}{rgb}{0,0,0.75}
\definecolor{mblue}{rgb}{0,0,0.75}
\definecolor{dyellow}{rgb}{0.5,0.5,0}
\definecolor{dmagenta}{rgb}{0.75,0,0.75}
\definecolor{dcyan}{rgb}{0,0.5,0.5}

\title{\huge Perfect Omniscience, Perfect Secrecy and Steiner Tree Packing}

\author{
Sirin Nitinawarat, {\em Student Member, IEEE}, and Prakash Narayan, {\em Fellow, IEEE}\\
\thanks{The work of S. Nitinawarat and P. Narayan was supported
by the National Science Foundation under Grants CCF0515124,
CCF0635271, CCF0830697 and InterDigital.  The material in this
paper was presented in part at the IEEE International Symposia on
Information Theory, Toronto, Ontario, Canada, July 2008, and
Seoul, Korea, June 2009.}
\thanks{The authors are with the Department of Electrical and
Computer Engineering and the Institute for Systems Research,
University of Maryland, College Park, MD 20742, USA.}
\noindent
\thanks{\hspace{-0.15in} Email: \{nitinawa, prakash\}@umd.edu} }

\begin{document}
\maketitle

\begin{abstract}

We consider perfect secret key generation for a ``pairwise
independent network'' model in which every pair of terminals share
a random binary string, with the strings shared by distinct
terminal pairs being mutually independent.  The terminals are then
allowed to communicate interactively over a public noiseless
channel of unlimited capacity.  All the terminals as well as an
eavesdropper observe this communication. The objective is to
generate a perfect secret key shared by a given set of terminals
at the largest rate possible, and concealed from the eavesdropper.

First, we show how the notion of perfect omniscience plays a central role in
characterizing perfect secret key capacity.  Second, a multigraph
representation of the underlying secrecy model leads us to an efficient
algorithm for perfect secret key  generation based on maximal Steiner tree
packing.  This algorithm attains capacity when all the
terminals seek to share a key, and, in general,
attains at least half the capacity.  Third, when a single
``helper'' terminal assists the remaining ``user'' terminals in
generating a perfect secret key, we give necessary and sufficient conditions
for the optimality of the algorithm; also, a ``weak'' helper
is shown to be sufficient for optimality.


\end{abstract}

\section{Introduction}

Given a collection of terminals $\mathcal{M}=\{1, \ldots, m\}$,
suppose that every pair $i, j$ of terminals, $1 \leq i < j \leq
m$, share a random binary string of length $e_{ij}$ (bits), with
the strings shared by distinct pairs of terminals being mutually
independent.  Then all the terminals are allowed to communicate
interactively in multiple rounds over a public noiseless channel
of unlimited capacity, with all such communication being observed
by all the terminals.  The main goal is to generate, for a given subset $A$
of the terminals in $\mathcal{M}$,
a {\em perfect secret key} (SK) namely shared uniformly distributed random bits
-- of the largest size -- such that these shared bits are exactly independent
of an eavesdropper's observations of the interterminal
communication.  All the terminals in $\mathcal{M}$ cooperate in
generating such a perfect SK for $A$.

This model for perfect SK generation, hereafter referred to as a
``pairwise independent network'' (PIN) model, is a specialized
version of an earlier PIN model \cite{YeRezSha06, YeRez07,
Nitinawarat_etal_IT}.  In the latter, every pair of terminals
observe a pair of correlated signals (not necessarily identical
as here) that are independent of pairs of signals observed by all
other terminal pairs.  In \cite{Nitinawarat_etal_IT}, we had
studied Shannon theoretic SK generation (not in the perfect sense)
in the asymptotic limit of large signal observation lengths, and
its connection to the combinatorial problem of Steiner tree
packing of a multigraph. Leading work on Shannon theoretic
SK generation with public communication originated in
\cite{Mau90, Mau93, AhlCsi93};  see also
\cite{CsiNar00} for related models.

In contrast with \cite{Nitinawarat_etal_IT},
the present work bears the essence of ``zero-error information
theory,'' and accordingly, we rely on mathematical techniques of
a combinatorial nature.  Specifically, our emphasis here is on
{\em perfect} SK generation for fixed signal observation lengths as well
as for their asymptotic limits.  For convenience, we
shall continue to refer to our present model as the PIN model. This
model possesses the appropriate structure for investigating
the concept of perfect SK in which the generated key is exactly
recoverable by every terminal in the secrecy seeking set $A$; is
exactly independent of the eavesdropper's observations; and is
uniformly distributed.  Also, its special structure makes for a
new concept of perfect omniscience, which plays a central role.
Furthermore, in the spirit of
\cite{Nitinawarat_etal_IT}, the PIN model reveals points of
contact between perfect SK generation and the combinatorial
problem of maximal Steiner tree packing of a multigraph.  We
remark that tree packing has been used in the context of network
coding (see, for instance \cite{Li_CISS04, WuJianKung_IT06}).

Our three main contributions described below are motivated by a
known general connection between (not
necessarily perfect) SK generation at the maximum rate
and the minimum communication for (not necessarily
perfect) omniscience \cite{CsiNar04, CsiNar08}, and
by the mentioned connection between the
former and the combinatorial problem of maximal Steiner tree
packing of a multigraph \cite{Nitinawarat_etal_IT}.

First, the concept of perfect omniscience enables us to obtain a
single-letter formula for the perfect SK capacity of the PIN
model; moreover, this capacity is shown to be achieved
by linear noninteractive communication, and coincides with the
(standard) SK capacity derived in our previous work
\cite{Nitinawarat_etal_IT}. This result establishes a connection
between perfect SK capacity and the minimum rate of communication
for perfect omniscience, thereby particularizing to the PIN model
a known general link between these notions {\em sans} the
requirement of the omniscience or secrecy being perfect
\cite{CsiNar04}.

Second, the PIN model can be represented by a multigraph.  Taking
advantage of this representation, we put forth an efficient
algorithm for perfect SK generation using a maximal packing of
Steiner trees of the multigraph.  This algorithm involves public
communication that is linear as well as noninteractive, and
produces a perfect SK of length equal to the maximum size of such
Steiner tree packing.  When all the terminals in $\mathcal{M}$
seek to share a perfect SK, the algorithm is shown to achieve
perfect SK capacity.  However, when only a subset of terminals in
$A \subset \mathcal{M}$ wish to share a perfect SK, the algorithm
can fall short of achieving capacity; nonetheless, it is shown to
achieve at least half of it. Additionally, we obtain nonasymptotic
and asymptotic bounds on the size and rate of the best perfect SKs
generated by the algorithm. {\em These bounds are of independent
interest from a purely graph theoretic viewpoint as they
constitute new estimates for the maximum size and rate of Steiner
tree packing of a given multigraph}.

Third, a special configuration of the PIN model arises when a lone
``helper'' terminal $m$ aids the ``user'' terminals in $A =
\mathcal{M} \backslash \{m\}$ generate a perfect SK.  This model
has two special features: firstly, (a single) terminal $m$
possesses all the bit strings that are not in $A$; secondly, a
Steiner tree for $A$ is a spanning tree for either $A$ or
$\mathcal{M}$.  These features enable us to obtain necessary and
sufficient conditions for Steiner tree packing to achieve perfect
SK capacity, as also a further sufficient condition that posits a
``weak'' role for the helper terminal $m$.

Preliminaries and the problem formulation are in Section II.  Our
results are described in Section III and proved in Section IV.  A
discussion follows in Section V.

\section{Preliminaries}

Suppose that the terminals in $\mathcal{M} = \{1, \ldots, m\},\ m
\geq 2,$ observe, respectively, $n$ independent and identically
distributed (i.i.d.) repetitions of the rvs $\tilde{X}_1, \ldots,
\tilde{X}_m,$ denoted by $\tilde{X}_1^n, \ldots, \tilde{X}_m^n,$
where $\tilde{X}_i^n = \left (\tilde{X}_{i,1}, \ldots,
\tilde{X}_{i,n} \right),~i \in \mathcal{M}$. We shall be concerned
throughout with a PIN model $\tilde{X}_1, \ldots, \tilde{X}_m$
\cite{YeRez07}, defined by each rv $\tilde{X}_i,~i \in
\mathcal{M},$ being of the form $\tilde{X}_i = \left( X_{ij},~j
\in \mathcal{M} \backslash \{i\} \right)$ with $m-1$ components,
and the ``reciprocal pairs'' of rvs $\{\left ( X_{ij}, X_{ji}
\right ),~1 \leq i < j \leq m \}$ being mutually independent.  We
assume further that $X_{ij}=X_{ji},~1 \leq i \neq j \leq m,$ where
$X_{ij}$ is uniformly distributed over the set of all binary
strings of length $e_{i j}$ (bits).  Thus, every pair of terminals
is associated with a random binary string that is independent of all other
random binary strings
associated with all other pairs of terminals.  The assumption is
tantamount to every pair of terminals $i, j$ sharing at the outset
privileged and pairwise ``perfect secrecy'' of $e_{i j}$ bits.
Following their observation of the random sequences as above, the
terminals in $\mathcal{M}$ are allowed to communicate among
themselves over a public noiseless channel of unlimited capacity;
all such public communication, which maybe interactive and
conducted in multiple rounds, is observed by all the terminals. A
communication from a terminal, in general, can be any function of
its observed sequence as well as all previous public
communication. The public communication of all the terminals will
be denoted collectively by $\mathbf{F} = \mathbf{F}^{(n)}$.

\textbf{ Definition 1:} The communication ${\bf F}$ is termed {\em
linear noninteractive communication} (LC) if ${\bf F} = (F_1,
\ldots, F_m)$ with\footnote{All additions and multiplications are
modulo 2.} $F_i = L_i \tilde{X}_i^n$, where $L_i$ is a $b_i \times
\big ( \sum_{j\,\neq\,i}\,n\,e_{i j} \big )$ matrix\footnote{It is
assumed that $\sum_{j\,\neq\,i}\,e_{i j} \geq 1,\  \
i=1, \ldots, m$.} with $\{0,1\}$-valued entries, $i=1, \ldots, m$.
The integer $b_i \geq 0,~i=1, \ldots, m$, represents the length
(in bits) of the communication $F_i$ from terminal $i$; the
overall communication ${\bf F}$ has length $\sum_{i\,=\,1}^m\,b_i$
(bits).

The primary goal is to generate shared perfect secret common randomness
for a given set $A \subseteq \mathcal{M}$ of terminals at the
largest rate possible, with the remaining terminals (if any)
cooperating in secrecy generation. The resulting perfect secret key must
be accessible to every terminal in $A$; but it need not be
accessible to the terminals not in $A$ and nor does it need to be
concealed from them. It must, of course, be kept perfectly secret from the
eavesdropper that has access to the public interterminal
communication $\mathbf{F}$, but is otherwise passive, i.e., unable
to tamper with this communication.

The following basic concepts and definitions are adapted
from \cite{CsiNar04, CsiNar08}.  For rvs $U, V$, we say
that $U$ is {\em perfectly recoverable} from $V$ if
$Pr\{U = f(V)\} = 1$ for some function $f(V)$.
With the rvs $K$ and $\mathbf{F}$ representing a secret key and
the eavesdropper's knowledge, respectively, information
theoretic {\em perfect secrecy} entails that the security
index\footnote{All logarithms are to the base 2.}
\begin{eqnarray}
    s(K;\mathbf{F})
    &=& \log{|\mathcal{K}|} - H(K) + I(K \wedge \mathbf{F})    \nonumber \\
    &=& \log{|\mathcal{K}|} - H(K|\mathbf{F})\ =\ 0, \label{eqn:Sec2-1}
\end{eqnarray}
 where $\mathcal{K}$ is the range of
$K$ and $|\centerdot|$ denotes cardinality. This requirement
simultaneously renders $K$ to be uniformly distributed
and independent of $\mathbf{F}$.

\textbf{ Definition 2:} Given any set $A \subseteq \mathcal{M}$ of
size $|A| \geq 2,$ a rv $K$ is a {\em perfect secret key} (SK) for
the set of terminals $A$ achievable with communication
$\mathbf{F}$, if $K$ is perfectly recoverable\footnote{ The extra
requirement of perfectness in recoverability is not a limiting
factor for the PIN model in contrast with other models of SK
generation.} from $\left (\tilde{X}_i^n, \mathbf{F}\right )$ for
each $i \in A$ and, in addition, it satisfies the perfect secrecy
condition (\ref{eqn:Sec2-1}).

\textbf{ Definition 3:} A number $R$ is an {\em achievable perfect
SK rate} for a set of terminals $A \subseteq \mathcal{M}$ if there
exist perfect SKs $K^{(n)}$ for $A$ achievable with appropriate
communication, such that
\begin{equation}
    \frac{1}{n}\log{|\mathcal{K}^{(n)}|} \rightarrow R\ \ \
    \mbox{as}\ \ \ n \rightarrow \infty,\nonumber
\end{equation}
where $\mathcal{K}^{(n)}$ is the range of $K^{(n)}$.  The largest
achievable perfect SK rate is the perfect SK capacity $C(A)$.

Thus, by definition, the perfect SK capacity for $A$ is the
largest rate of a rv that is perfectly recoverable at each
terminal in $A$ from the aggregate information available to it,
and is uniformly distributed and concealed from an eavesdropper
with access to the public interterminal communication; it need not
be concealed from the terminals in $A^c = \mathcal{M} \backslash
A$, which cooperate in secrecy generation. The notion of perfect
SK capacity is more stringent than that of SK capacity under the
requirements of the key being asymptotically recoverable for each
$i \in A$ and the security index tending to $0$, both as $n
\rightarrow \infty$; in particular, now the security index must
equal zero for all sufficiently large $n$. The latter SK capacity
for the PIN model has been characterized in
\cite{Nitinawarat_ISIT08, Nitinawarat_Allerton08,
Nitinawarat_etal_IT}.

A central role is played by the notion of {\em perfect
omniscience} which is a strict version of the concept of {\em
omniscience} introduced in \cite{CsiNar04}. {\em This notion
does not involve any secrecy requirements}.

\textbf{ Definition 4:} The communication ${\bf F}$ is {\em
communication for perfect omniscience} for $A$ if $(\tilde{X}_1^n,
\ldots, \tilde{X}_m^n)$ is perfectly recoverable from
$(\tilde{X}_i^n, {\bf F})$ for every $i \in A$.  Further, ${\bf
F}$ is {\em linear noninteractive communication for perfect
omniscience} ($\mbox{LCO}^{(n)}(A)$) if ${\bf F}$ is an LC and
satisfies the previous perfect recoverability condition.  The
minimum length (in bits) of an $\mbox{LCO}^{(n)}(A)$, i.e.,
$\min_{\mbox{LCO}^{(n)}(A)} \sum_{i\,=\,1}^m\,b_i$, will be
denoted by $\mbox{LCO}^{(n)}_m(A)$.  The {\em minimum rate} of
$\mbox{LCO}^{(n)}(A)$ is $OMN(A) \triangleq \limsup_n \frac{1}{n}
\mbox{LCO}_{m}^{(n)}(A).$

\section{Results}

\subsection{Perfect SK Capacity for the PIN Model}

Our first main contribution is a (single-letter) characterization of
the perfect SK capacity for the PIN model, which brings forth a connection
with the minimum rate of communication for perfect omniscience.

\vspace{0.1in}
{\em \textbf{ Theorem 1:} The perfect SK capacity
for a set of terminals $A \subseteq \mathcal{M}$ is
\begin{equation}
   C(A)~ = ~ \sum_{i, j}\,e_{i j} \,-\, OMN(A)  \label{eqn:Sec3.1-1}
\end{equation}
where
\begin{equation}
    OMN(A) ~=~ \min_{(R_1, \ldots, R_m)~\in~\mathcal{R}(A)}\
          \sum_{i\,=\,1}^m\,R_i,    \label{eqn:Sec3.1-2}
\end{equation}
with
\vspace{0.05in}
\[\hspace{-2.7in} \mathcal{R}(A) =\]
\begin{equation}
\left \{
\begin{array}{c}
    (R_1, \ldots, R_m)\,\in\,\mathbb{R}^m: \ R_i \geq 0,\ i=1, \ldots, m, \\
    \sum_{i\,\in\,B} \,R_i ~\geq~ \sum_{1 \leq i < j \leq m,\ i \in B,\ j \in B} ~e_{i j}, \\
    \forall B \nsupseteq A,\ \emptyset \neq B \subset \mathcal{M}
\end{array}
\right \}.  \label{eqn:Sec3.1-3}
\end{equation}
Furthermore, this perfect SK capacity can be achieved with linear
noninteractive communication.}

{\em Remarks:} (i) Clearly, the perfect SK capacity, by definition,
cannot exceed the (standard) SK capacity studied in
\cite{Nitinawarat_ISIT08, Nitinawarat_etal_IT}.  Indeed, Theorem 1
implies that the latter is attained by a perfect SK.

(ii) In the same vein, the minimum rate of communication for
(asymptotic) omniscience \cite{CsiNar04} can be attained for the
PIN model with perfect recoverability  at $A$ of $(\tilde{X}_1^n,
\ldots, \tilde{X}_m^n)$ for all $n$ sufficiently large, and with
linear noninteractive communication.  We mention that
noninteractive communication, without a claim of linearity, was
shown to suffice for (asymptotic) omniscience in \cite{CsiNar04}.

\subsection{Maximal Steiner Tree Packing and Perfect SK Generation}

Theorem 1 serves to establish the sufficiency of an LC in
achieving perfect SK capacity through the intermediate attainment
of perfect omniscience for $A$, as seen in its proof below.
However, as also evident from the proof, decoding is by exhaustive
search of prohibitive complexity.

The PIN model can be represented by a multigraph.  This
representation leads us to an efficient algorithm for perfect SK
generation, not necessarily through perfect omniscience, by a maximal packing of
Steiner trees of the multigraph.  In particular, this algorithm
will be seen to entail public communication in the form of an LC.
On the other hand, such an algorithm based on maximal Steiner tree
packing need not attain perfect SK capacity.  The size of the
largest perfect SK that is thus generated can be estimated in
terms of the minimum length of an $\mbox{LCO}^{(n)}(A)$.

\textbf{Definition 5:} A {\em multigraph} $~G=\left (V, E \right)$
with vertex set $V$ and edge set $E$ is a connected undirected
graph with no selfloops and with multiple edges possible between
any pair of vertices.  Given $G=\left (V, E \right)$ and a
positive integer $n$, let $G^{(n)}=\left (V, E^{(n)} \right )$
denote the multigraph with vertex set $V$ and edge set $E^{(n)}$
wherein every vertex pair is connected by $n$ times as many edges
as in $E$; in particular, $G^{(1)}=G$.  Furthermore, $|E^{(n)}|$
will denote the total number of edges in $E^{(n)}$.

To the PIN model $\tilde{X}_1, \ldots, \tilde{X}_m$ (cf. section II),
we can associate a multigraph $G=(\mathcal{M}, E)$
with $\mathcal{M} = \{1, \ldots, m\}$ and the number of edges
connecting a vertex pair $(i, j)$ in $E$ equal to $e_{ij}$; in
particular, the edge connecting $(i,j)$ will be associated with
the random binary string $X_{ij}$.

By this association, it will be convenient to represent
(\ref{eqn:Sec3.1-2}) and (\ref{eqn:Sec3.1-3}) as
\begin{equation}
    OMN_G(A) \ = \ \min_{(R_1, \ldots, R_m)~\in~\mathcal{R}_G(A)}\
    \sum_{i\,=\,1}^m\,R_i,    \label{eqn:Sec3.2-4}
\end{equation}
with
\[\hspace{-2.5in} \mathcal{R}_G(A) =\]
\begin{equation}
\left \{
\begin{array}{l}
    \vspace{0.05in} (R_1, \ldots, R_m)\,\in\,\mathbb{R}^m: \ R_i \geq 0,~i=1, \ldots, m, \\
    \sum_{i\,\in\,B}\,R_i \geq ~ \sum_{1 \leq i < j \leq m,\ i \in B,\ j \in B} ~e_{ij}, \\
    \vspace{0.05in} \ \ \ \ \ \ \ \ \ \ \ \forall B \nsupseteq A,\ \emptyset \neq B \subset \mathcal{M}
\end{array}
\right \}, \label{eqn:Sec3.2-5}
\end{equation}
whereupon (\ref{eqn:Sec3.1-1}) can be restated as
\begin{equation}
    C(A) \ =\  |E| - OMN_G(A).   \label{eqn:Sec3.2-6}
\end{equation}
Furthermore, it is easy and useful to note that for every $n \geq
1$,
\begin{equation}
    OMN_{G^{(n)}}(A) \ =\ n OMN_G(A). \label{eqn:Sec3.2-7}
\end{equation}

\textbf{Definition 6:} For $A \subseteq V$, a {\em Steiner tree}
(for $A$) of $G=(V, E)$ is a subgraph of $G$ that is a tree, i.e.,
containing no cycle, and whose vertex set contains $A$; such a
Steiner tree is said to {\em cover} $A$.  A {\em Steiner tree
packing} of $G$ is any collection of edge-disjoint Steiner trees
of $G$. Let $\mu(A, G)$ denote the {\em maximum} size of such a
packing (cf. \cite{Die05}), i.e., the maximum number of trees in
the packing.  The {\em maximum rate}\footnote{In fact, $\lim_{n
\rightarrow \infty} \frac{1}{n} \mu(A, G^{(n)})$ exists, as shown
later in Proposition 4.} of Steiner tree packing of $G$ is
$\limsup_{n \rightarrow \infty} \frac{1}{n} \mu(A, G^{(n)})$. When
$A = V$, a Steiner tree becomes a {\em spanning tree}, with
corresponding notions of {\em spanning tree packing}, maximum size
and rate.

Given a PIN model, the notion of Steiner tree packing of the
associated multigraph leads to an efficient algorithm for
constructing an $\mbox{LCO}^{(n)}(A)$ and thereby generating a
perfect SK. The next Theorem 2 indicates that the largest size of
a perfect SK that the algorithm generates is the maximum size of
the Steiner tree packing.  Furthermore, Theorem 2 and its
corollary, and Theorem 5 provide nonasymptotic and asymptotic
bounds on the size and rate, respectively, of the best perfect SKs
generated by the algorithm. {\em Of independent interest from a
purely graph theoretic viewpoint, these results also constitute
new bounds for the maximum size and rate of Steiner tree packing
of a given multigraph.}

\vspace{0.1in}
{\em \textbf{ Theorem 2:}  For the multigraph
$G=(\mathcal{M}, E)$ associated with a PIN model and for $A
\subseteq \mathcal{M}$, it holds for every $n \geq 1$ that

{\bf(i)} the terminals in $\mathcal{M}$ can devise an
$\mbox{LCO}^{(n)}(A)$ of total length $~n|E^{(1)}| - \mu(A,
G^{(n)})~$ and subsequently generate a perfect SK $~K^{(n)}$ with
$\log{|\mathcal{K}^{(n)}|} = \mu(A, G^{(n)})$;

\begin{equation} \hspace{-0.8in} \mbox{{\bf(ii)}}~ \mu(A, G^{(n)}) ~\leq~ n |E^{(1)}| -
\mbox{LCO}_{m}^{(n)}(A);  \label{eqn:Sec3.2-7-2}
\end{equation}

{\bf(iii)} furthermore, $\mbox{LCO}_{m}^{(n)}(A)$ is bounded below
by the value of an integer linear program according to
\begin{equation}
    \mbox{LCO}_{m}^{(n)}(A) ~\geq~ INT_{G^{(n)}}(A)  \nonumber
\end{equation}
where
\begin{equation}
    INT_{G^{(n)}}(A) ~=~ \min_{(I_1, \ldots, I_m)~\in~\mathcal{I}_{G^{(n)}}(A)}\
          \sum_{i\,=\,1}^m\,I_i,               \label{eqn:Sec3.2-8}
\end{equation}
with
\vspace{0.1in}
\[\hspace{-2.3in} \mathcal{I}_{G^{(n)}}(A) ~=~ \]
\begin{equation}
\left \{
\begin{array}{c}
    (I_1, \ldots, I_m)\,\in\,\mathbb{Z}^m: \ I_i \geq 0,\ i=1, \ldots, m, \\
    \sum_{i \in B}\,I_i ~\geq~ n~\sum_{1 \leq i < j \leq m,\ i \in B,\ j \in B} ~e_{ij}, \\
    \forall B \nsupseteq A,\ \emptyset \neq B \subset \mathcal{M}
\end{array}
\right \}.  \label{eqn:Sec3.2-9}
\end{equation}}

\vspace{0.1in} {\em \textbf{Corollary 3:} For every $n \geq 1$,
the maximum size of Steiner tree packing of a multigraph $G^{(n)}$
satisfies
\begin{equation}
    \mu(A, G^{(n)}) ~\leq~ n~|E^{(1)}| - INT_{G^{(n)}}(A), \label{eqn:Sec3.2-10}
\end{equation}
with equality when $A = \mathcal{M}$.}

{\em Remarks:}  (i) Note that the bounds in Theorem 2 are
nonasymptotic, i.e., valid for every $n$.  Also, note in the bound
in Theorem 2 (ii) for $\mu(A,G^{(n)})$ that
$\mbox{LCO}_{m}^{(n)}(A)$ is defined in terms of its {\em
operational significance.}

(ii) Further, Theorem 2 provides a nonasymptotic {\em computable}
lower bound for $\mbox{LCO}_{m}^{(n)}(A)$ in terms of an integer
linear program.  The optimum value of its linear programming
relaxation constitutes a further lower bound which equals
$OMN_{G^{(n)}}(A) = n OMN_G(A),$ by (\ref{eqn:Sec3.2-7}).
\\

Next, we turn to connections between perfect SK capacity
$C(A)$ and the maximum rate of Steiner tree packing of
$G=(\mathcal{M}, E)$. The following concept of ``fractional''
Steiner tree packing will be relevant.

For $A \subseteq \mathcal{M} = \{1, \ldots, m\}$, consider the
collection $\{S_1, \ldots, S_{k}\}$ of {\em all distinct}
Steiner trees (for $A$) of $G$, where $k=k(G)$.  Consider the region
\[\hspace{-2.6in} \mathcal{T}_G(A) =\]
\begin{equation}
\left \{
\begin{array}{c}
    (T_1, \ldots, T_k)\,\in\,\mathbb{R}^k: \ T_l \geq 0,\ l=1, \ldots, k, \\
    \sum_{l: (i,j)~\in~S_l}\,T_l \leq ~e_{ij}\\
    \forall (i, j),\ 1 \leq i < j \leq m
\end{array}
\right \}. \label{eqn:Sec3.2-11}
\end{equation}

\textbf{Definition 7:} For a multigraph $G = (\mathcal{M}, E)$ and
$A \subseteq \mathcal{M}$, the {\em maximal ``fractional'' Steiner
tree packing} of $G$, denoted $\mu_f(A, G)$, is $\mu_f(A, G)
\triangleq \max_{\mathcal{T}_G(A)}
    \sum_{l = 1}^k  T_l.$

{\em Remarks:} (i) Clearly, $\mu_f(A,G)$ corresponds to a linear
program with finite optimum value, and the maximum is attained.
Furthermore, it is readily verified that for every $n \geq 1$,
\begin{equation}
    \mu_f(A, G^{(n)})\ =\ n\,\mu_f(A, G).  \label{eqn:Sec3.2-12}
\end{equation}

(ii) We observe that in Definition 6, $\mu(A, G) \triangleq
\max_{\mathcal{T}_G(A) \cap \mathbb{Z}^k}
    \sum_{l = 1}^k  T_l.$

\vspace{0.1in}
{\em \textbf{Proposition 4:} For a multigraph
$G=(\mathcal{M}, E)$ and $A \subseteq \mathcal{M}$, it holds that
the maximum rate of Steiner tree packing (for $A$) of $G$
satisfies
\begin{eqnarray}
    \limsup_{n \rightarrow \infty} \frac{1}{n} \mu(A, G^{(n)}) &=&
    \liminf_{n \rightarrow \infty} \frac{1}{n} \mu(A, G^{(n)}) \nonumber \\
    &=&
    \lim_{n \rightarrow \infty} \frac{1}{n} \mu(A, G^{(n)}) \nonumber \\
    &=& \mu_f(A,G).    \label{eqn:Sec3.2-13}
\end{eqnarray}}

\vspace{0.1in}
{\em \textbf{Theorem 5:} For the multigraph
$G=(\mathcal{M}, E)$ associated with the PIN model and for $A
\subseteq \mathcal{M}$, it holds that
\begin{equation}
    \frac{1}{2}C(A) ~\leq~ \lim_{n \rightarrow \infty} \frac{1}{n} \mu(A, G^{(n)}) ~\leq~ C(A).
    \label{eqn:Sec3.2-14}
\end{equation}
Furthermore, when $A = \mathcal{M}$,
\begin{equation}
    \lim_{n \rightarrow \infty} \frac{1}{n} \mu(\mathcal{M}, G^{(n)}) ~=~ C(\mathcal{M}).
    \label{eqn:Sec3.2-14-b}
\end{equation}}

{\em Remark:}  For the PIN model with $m$ terminals, every Steiner
tree has at most $m-1$ edges.  Also, from (\ref{eqn:Sec3.2-14}), $
\mu(A, G^{(n)}) \lesssim n C(A)$ for all large $n$. Hence, the
overall complexity of the perfect SK generation algorithm based on
Steiner tree packing is linear (in $n$).\newline

The upper bound on $\lim_{n \rightarrow \infty} \frac{1}{n} \mu(A,
G^{(n)})$ in Theorem 5 is not tight, in general, as seen by the
following example.

\textbf{Example:}  Consider the multigraph \cite{Li_CISS04} in
Figure 1 with $|\mathcal{M}| = 7$ and $|A|=4$; the terminals in
$A$ are represented by the solid circles and every shown edge is
single. Computations give that $C(A) = 2.0$ by
(\ref{eqn:Sec3.2-6}), (\ref{eqn:Sec3.2-4}), while $\lim_{n
\rightarrow \infty} \frac{1}{n}\mu(G^{(n)},A) = 1.8$ by
Proposition 4 and the scheme in Lemma 1.

\vspace{-0.05in}
\begin{figure}[h]
\begin{center}
\setlength{\unitlength}{1bp}%
\begin{picture}(200, 80)(0,0)
     \put(100,80){\circle*{7}}
     \put(100,80) {\line(1,-1){30}}
     \put(70,50){\circle{7}}
     \put(100,80) {\line(0,-1){30}}
     \put(100,50){\circle{7}}
     \put(100,80) {\line(-1,-1){30}}
     \put(130,50){\circle{7}}

     \put(70,20) {\line(0,1){30}}
     \put(70,20) {\line(1,1){30}}
     \put(70,20){\circle*{7}}

     \put(100,20) {\line(-1,1){30}}
     \put(100,20) {\line(1,1){30}}
     \put(100,20){\circle*{7}}

     \put(130,20) {\line(0,1){30}}
     \put(130,20) {\line(-1,1){30}}
     \put(130,20){\circle*{7}}
\end{picture}
\end{center}
\end{figure}
\vspace{-0.3in} $\hspace{1.0in} \mbox{Figure 1: Example}$

\subsection{The Single Helper Case}

As observed after Theorem 5, the maximum rate of Steiner tree
packing can fail to achieve perfect SK capacity.  A natural
question that remains open is whether the maximum rate of Steiner
tree packing equals perfect SK capacity for the special case of
the PIN model in which a lone ``helper'' terminal $m$ assists the
``user'' terminals in $A = \{1, \ldots, m-1\}$ generate a perfect
SK.  In this section, we provide partial answers.

First, we derive necessary and sufficient conditions for the
maximum rate of Steiner tree packing to equal perfect SK capacity
in (\ref{eqn:Sec3.2-14}) and, analogously, the (nonasymptotic)
maximum size of Steiner tree packing to meet its upper bound in
(\ref{eqn:Sec3.2-10}).  These conditions entail the notion of a
{\em fractional} multigraph. Throughout this section, we shall
assume that $A = \{1, \ldots, m-1\} \subset \mathcal{M} = \{1,
\ldots, m\}$.

\textbf{Definition 8:}  Given a multigraph $G=(\mathcal{M}, E)$ as
in Definition 5, a {\em fractional multigraph} $\tilde{G}=(A,
\tilde{E})$ in $A$ (with vertex set $A$) has edge set $\tilde{E} =
\{ \tilde{e}_{ij} \in \mathbb{R},\ 0 \leq \tilde{e}_{ij} \leq
e_{ij},\ 1\leq i < j \leq m - 1\}$.  For any such $\tilde{G}$, the
{\em complementary fractional multigraph} $G \backslash \tilde{G}
= (\mathcal{M}, E \backslash \tilde{E})$ has vertex set
$\mathcal{M}$ and edge set $E \backslash \tilde{E} \triangleq
\{e_{ij} - \tilde{e}_{ij},\ 1 \leq i < j \leq m-1;\ e_{im},\ 1
\leq i \leq m-1\}$.  The definitions of $\mathcal{R}_G(A)$ in
(\ref{eqn:Sec3.2-5}), $OMN_G(A)$ in (\ref{eqn:Sec3.2-4}),
$\mathcal{T}_G(A)$ in (\ref{eqn:Sec3.2-11}) and $\mu_f(A,G)$ in
Definition 7 all have obvious extensions to $\tilde{G}$ and $G
\backslash \tilde{G}$ as well.  Further, (\ref{eqn:Sec3.2-7}) and
(\ref{eqn:Sec3.2-12}) also hold for $\tilde{G}$ and $G \backslash
\tilde{G}$.

\vspace{0.1in}
{\em \textbf{Proposition 6:}  For the multigraph $G
= (\mathcal{M}, E)$ associated with the PIN model, the following
hold:

{\bf (i)}
\begin{equation}
    \mu_f(A,G) \geq \max_{\tilde{G}} \mu_f(A, \tilde{G}) +
    \mu_f(\mathcal{M}, G \backslash \tilde{G}); \nonumber
\end{equation}

{\bf (ii)}
\begin{equation}
    {OMN}_G(A) \leq \min_{\tilde{G}} {OMN}_{\tilde{G}}(A) +
    {OMN}_{G \backslash \tilde{G}}(\mathcal{M}); \nonumber
\end{equation}

{\bf (iii)}
\begin{equation}
    \mu(A,G) \geq \max_{\tilde{G}_I} \mu(A, \tilde{G}_I) +
    \mu(\mathcal{M}, G \backslash \tilde{G}_I); \nonumber
\end{equation}

{\bf (iv)}
\begin{equation}
    {INT}_G(A) \leq \min_{\tilde{G}_I} {INT}_{\tilde{G}_I}(A) +
    {INT}_{G \backslash \tilde{G}_I}(\mathcal{M}), \nonumber
\end{equation}
where the optima in (i) and (ii) are over all fractional
multigraphs $\tilde{G}=(A, \tilde{E})$ in $A$, and the optima in
(iii) and (iv) are over all multigraphs $\tilde{G}_I = (A,
\tilde{E})$ in $A$ for which $\tilde{E}$ consists of only
integer-valued $\tilde{e}_{ij}$s.}

\vspace{0.1in} {\em \textbf{Theorem 7:} For the multigraph
$G=(\mathcal{M}, E)$ associated with the PIN model,

{\bf (i)}
\begin{equation}
    \lim_{n \rightarrow \infty}
    \frac{1}{n}\mu(A, G^{(n)})
     \ =\  C(A) \label{eqn:Sec3.3-15}
\end{equation}
iff
\begin{equation}
    {OMN}_G(A) = \min_{\tilde{G}}~ {OMN}_{\tilde{G}}(A) +
{OMN}_{G \backslash \tilde{G}}(\mathcal{M}), \label{eqn:Sec3.3-16}
\end{equation}
where the minimum is over all fractional multigraphs
$\tilde{G}=(A, \tilde{E})$ in $A$;

{\bf (ii)}
\begin{equation}
    \mu(A, G^{(n)})
     = |E| - {INT}_G(A) \nonumber
\end{equation}
iff
\begin{equation}
    {INT}_G(A) = \min_{\tilde{G}_I}~ {INT}_{\tilde{G}_I}(A) +
    {INT}_{G \backslash \tilde{G}_I}(\mathcal{M}), \label{eqn:Sec3.3-17}
\end{equation}
where the minimum is over all multigraphs $\tilde{G}_I=(A,
\tilde{E})$ for which $\tilde{E}$ consists of only integer-valued
$\tilde{e}_{ij}$s.}

\vspace{0.1in} Our final result provides another sufficient
condition for the maximum rate of Steiner tree packing to equal
perfect SK capacity. Recall from Theorem 1 that, in general,
perfect SK capacity for $A$ can be attained with public
communication that corresponds to the minimum communication for
perfect omniscience. If the latter can be accomplished with the
sole helper terminal $m$ communicating ``sparingly,'' then it
transpires that maximal Steiner tree packing attains the best
perfect SK rate.  An analogous nonasymptotic version of this claim
also holds. Heuristically, a sufficient ``weak'' role of the
helper terminal $m$ turns the Steiner tree packing of $A$, in
effect, into a spanning tree packing of $A$.

Let $d_i \triangleq \sum_{j \neq i} e_{ij}$ denote the degree of
vertex $i,\ i \in \mathcal{M}$.  Clearly, any $(R_1^*, \ldots,
R_m^*)$ (resp. $(I_1^*, \ldots, I_m^*)$) that attains the minimum
corresponding to $OMN_{G}(A)$ (cf. (\ref{eqn:Sec3.2-4})) (resp.
$INT_{G}(A)$ (cf. (\ref{eqn:Sec3.2-8}))) must satisfy $R_i^* \leq
d_i$ (resp. $I_i^* \leq d_i$), $i = 1, \ldots, m$.

\vspace{0.1in}
{\em \textbf{Theorem 8:} For the multigraph
$G=(\mathcal{M}, E)$ associated with the PIN model,

{\bf (i)} if there exists $(R_1^*, \ldots, R_m^*)$ that attains
$OMN_G(A)$ (cf. (\ref{eqn:Sec3.2-4})) with $R_m^* \leq d_m/2$,
then
\[\lim_{n \rightarrow \infty} \frac{1}{n}\mu(A, G^{(n)})\ =\
C(A)\ =\ |E| - OMN_G(A). \]

{\bf (ii)} if there exists $(I_1^*, \ldots, I_m^*)$ that attains
$INT_G(A)$ (cf. (\ref{eqn:Sec3.2-8})) with $I_m^* \leq \lfloor
d_m/2 \rfloor$, then
\[\mu(A, G)\ =\ |E| - INT_{G}(A). \]
}

\section{Proofs}

\textbf{Proof of Theorem 1:}  From remark (i) following Theorem 1,
we need prove only the achievability part.  The main step is to
show, using a random coding argument, the existence with large
probability of an $\mbox{LCO}^{(n)}(A)$ of small length under
appropriate conditions; the terminals in $A$ then extract from the
corresponding perfect omniscience a perfect SK of optimum rate.

Let $\tilde{X}_{\mathcal{M}}^n = \left( \tilde{X}_1^n, \ldots,
\tilde{X}_m^n \right)$ take values in
$\tilde{\mathcal{X}}_{\mathcal{M}}^n = \tilde{\mathcal{X}}_1^n
\times \ldots \times \tilde{\mathcal{X}}_m^n$, where
$\tilde{\mathcal{X}}_i^n = \{0,1\}^{\sum_{j\,\neq\,i}\,n\,e_{i
j}}$. We denote a realization of $\tilde{X}_{\mathcal{M}}^n$ by
$\tilde{x}_{\mathcal{M}}^n = \left( \tilde{x}_1^n, \ldots,
\tilde{x}_m^n \right)$.  Fix $b_1, \ldots, b_m$.  Let $ {\bf L}=
\left(  {\bf L}_1, \ldots,  {\bf L}_m \right)$ consist of mutually
independent random matrices of appropriate dimensions as in
Definition $1$.  Furthermore, the rv $ {\bf L}_i$ consists of
i.i.d. equiprobable components, $i=1, \ldots, m$.  Clearly, $ {\bf
L}_1, \ldots,  {\bf L}_m$ makes for a random LC.

Since for ${\bf L}_1, \ldots, {\bf L}_m$ to constitute an
$\mbox{LCO}^{(n)}(A)$, it suffices that the mapping
\begin{equation}
\nonumber \tilde{x}_{\cal M}^n \ \ \rightarrow (\tilde{x}_i^n,
{\bf L}_1 \tilde{x}_1^n, \ldots, {\bf L}_m \tilde{x}_m^n)
\end{equation}
be one-to-one for every $i \in A$, we have
\[\hspace{-1.0in}
  Pr \{~{\bf L}~\mbox{does~not~constitute~an~}\mbox{LCO}^{(n)}(A)~\}
\]
\vspace{-0.2in}
\begin{eqnarray}
    &=& Pr \left \{
        \begin{array}{c}
            \exists~ \tilde{x}_{\mathcal{M}}^n \neq \tilde{y}_{\mathcal{M}}^n
                \in \tilde{\mathcal{X}}_{\mathcal{M}}^n\ \mbox{satisfying} \\
            \tilde{x}_j^n = \tilde{y}_j^n~\mbox{for~some~}j \in A\ \mbox{such~that} \\
            {\bf L}_i \tilde{x}_i^n = {\bf L}_i \tilde{y}_i^n\
            \mbox{for~each}\ i=1, \ldots, m
        \end{array}
        \right \} \nonumber \\
   \vspace{0.1in}
   &=& Pr \left \{
        \begin{array}{c}
            \exists~ \tilde{x}_{\mathcal{M}}^n \neq {\bf 0}
                \in \tilde{\mathcal{X}}_{\mathcal{M}}^n\ \mbox{satisfying} \\
            \tilde{x}_j^n = {\bf 0} \mbox{~for~some~}j \in A\ \mbox{such~that} \\
            {\bf L}_i \tilde{x}_i^n =  {\bf 0}\
            \mbox{for~each}\ i=1, \ldots, m
        \end{array}
       \right \}  \label{eqn:Sec4-1} \\
   \vspace{0.1in}
   &\leq& \hspace{-0.1in} \sum_{\substack{
                    B \neq \emptyset, \\
                    B\,\nsupseteq\,A
          }}
      Pr \left \{
        \begin{array}{c}
           \exists~ \tilde{x}_{\mathcal{M}}^n
                \in \tilde{\mathcal{X}}_{\mathcal{M}}^n\
                \mbox{satisfying} \\
            \tilde{x}_j^n \neq  {\bf 0}~\forall j \in B,~\mbox{and~}
                \tilde{x}_j^n = {\bf 0}~\forall j \in B^c\ \\
            \mbox{such~that}\ {\bf L}_i \tilde{x}_i^n = {\bf 0} \\
            \mbox{for~each}\ i=1, \ldots, m
        \end{array}
      \right \}, \nonumber \\ \label{eqn:Sec4-2}
\end{eqnarray}
where (\ref{eqn:Sec4-1}) is by the linearity of the communication
and (\ref{eqn:Sec4-2}) is obtained by applying the union bound to
the event in (\ref{eqn:Sec4-1}).

Now, we note by the assumed independence of ${\bf L}_1, \ldots
{\bf L}_m$ and the fact that the components of ${\bf L}_i$ are
i.i.d. and equiprobable, $i = 1, \ldots, m$, that for each
nonempty $B \nsupseteq A$, and any $\tilde{x}_{\mathcal{M}}^{n}$
satisfying $\tilde{x}_j^n \neq  {\bf 0}~\forall j \in B,~
\mbox{and~} \tilde{x}_j^n = {\bf 0}~\forall j \in B^c$, we have

\vspace{-0.05in}
\[\hspace{-1.0in} Pr \{{\bf L}_i \tilde{x}_i^n = {\bf 0}\ \mbox{for~every}\ i = 1, \ldots, m \}\]
\vspace{-0.25in}
\begin{eqnarray}
    &=& Pr \{{\bf L}_i \tilde{x}_i^n = {\bf 0}\ \mbox{for~every}\ i \in B \} \nonumber \\
    \vspace{0.3in} &=& \prod_{i\,\in\,B} 2^{- b_i} =
    2^{-\sum_{i\,\in\,B}\,b_i}. \label{eqn:Sec4-3}
\end{eqnarray}

Continuing with $(\ref{eqn:Sec4-2})$ upon using
$(\ref{eqn:Sec4-3})$, we obtain

\vspace{0.0in}
\[\hspace{-1.0in}
  Pr \{~{\bf L}~\mbox{does~not~constitute~an~}\mbox{LCO}^{(n)}(A)~\}
\]
\begin{eqnarray}
   \vspace{-0.05in}
   &\leq& \sum_{
                \substack{
                    B \neq \emptyset, \\
                    B\,\nsupseteq\,A
          }}
        \Biggl\lvert \Biggl\{
        \begin{array}{l}
            \tilde{x}_{\mathcal{M}}^n
            \in \tilde{\mathcal{X}}_{\mathcal{M}}^n :
            \tilde{x}_j^n \neq  {\bf 0} \\
            \forall j \in B,\,\tilde{x}_j^n = {\bf 0}~\forall j \in B^c
        \end{array}
        \Biggr\}
        \Biggr\rvert~2^{-\sum_{i\,\in\,B}\,b_i}
        \nonumber \\
   &\leq& \sum_{
                \substack{
                    B \neq \emptyset, \\
                    B\,\nsupseteq\,A
          }}
        2^{n \left( \sum_{l, k\,\in\,B}\,e_{l k} \right)}
            2^{- \sum_{i\,\in\,B}\,b_i} \nonumber \\
   &=&  \sum_{
                \substack{
                    B \neq \emptyset, \\
                    B\,\nsupseteq\,A
          }}
        2^{-n \left( \frac{1}{n} \sum_{i\,\in\,B}\,b_i
            - \sum_{l, k\,\in\,B}\,e_{l k} \right) }.  \label{eqn:Sec4-4}
\end{eqnarray}
We note that in this proof, the special structure of the PIN model
is used for the first time in the second inequality above.

Now, let $\left( R_1^*, \ldots, R_m^* \right)$ achieve the minimum
in the right-side of (\ref{eqn:Sec3.1-2}).  Pick an arbitrary
$\epsilon > 0$ and choose $b_i$ in $(\ref{eqn:Sec4-4})$ as $b_i =
\lceil n(R_i^* + \epsilon) \rceil,\ i = 1, \ldots, m$. Then, by
the definition of $\mathcal{R}(A)$, the right side of
(\ref{eqn:Sec4-4}) decays to zero exponentially rapidly in $n$; in
particular, we get from that for all $n$ sufficiently large, ${\bf
L}$ constitutes an $\mbox{LCO}^{(n)}(A)$ with large probability.
This implies the existence of a (deterministic) $L = \left( L_1,
\ldots, L_m \right)$ that constitutes an $\mbox{LCO}^{(n)}(A)$ for
all $n$ sufficiently large.

It remains to extract a perfect SK from the perfect omniscience
obtained above.  By the definition of the PIN model, observe that
\[ Pr\{\tilde{X}_{\mathcal{M}}^n = \tilde{x}_{\mathcal{M}}^n\} =
2^{-\sum_{l, k}\,n\,e_{l k}}~\mbox{for
all}~\tilde{x}_{\mathcal{M}}^n \in
\tilde{\mathcal{X}}_{\mathcal{M}}^n.\] By the linearity of the
$\mbox{LCO}^{(n)}(A)$ above, it is readily seen that \newline the
cardinality
$|\{\tilde{x}_{\mathcal{M}}^n\,\in\,\tilde{\mathcal{X}}_{\mathcal{M}}^n\,:\,L_i
\tilde{x}_i^n = a_i,\,i = 1, \ldots, m \}|$ is the same for all
feasible $\left( a_1, \ldots, a_m \right)$ where $a_i \in
\{0,1\}^{b_i},\ i = 1, \ldots, m,$ and that this common number is
at least \[N = 2^{ \left( \sum_{l, k}\,n\,e_{l k} \right) - \left(
\sum_{i\,=\,1}^m\,b_i \right)}.\]  For each communication message
$(a_1, \ldots, a_m)$, we index the elements of the {\em coset} $\{
\tilde{x}_{\mathcal{M}}^n: L_i \tilde{x}_i^n = a_i,\ i = 1,
\ldots, m \}$ in a fixed manner.  Then, for a realization
$\tilde{x}_{\mathcal{M}}^n \in
\tilde{\mathcal{X}}_{\mathcal{M}}^n$, every terminal in $A$ (which
knows $\tilde{x}_{\mathcal{M}}^n$ by omniscience) picks as the
perfect SK the index of $\tilde{x}_{\mathcal{M}}^n$ in its coset,
as in \cite{YeNar_ISIT05}.  Since $\tilde{X}_{\mathcal{M}}^n$
takes values in $\tilde{\mathcal{X}}_{\mathcal{M}}^n$ and since
each coset has the same size, it follows that this random index is
uniformly distributed and independent of the coset (the
communication message), thereby constituting a perfect SK. Lastly,
the rate of this perfect SK is at least
\begin{eqnarray}
    \lim_{n \rightarrow \infty} \frac{1}{n}\log{N}
    &=& \sum_{l, k}\,e_{l k} \, - \, \sum_{i\,=\,1}^m R_i^* - m\epsilon \nonumber \\
    &=& \sum_{l, k}\,e_{l k} \, - \, OMN(A) - m\epsilon, \nonumber
\end{eqnarray}
where $\epsilon > 0$ is arbitrary. $\ \ \blacksquare$

\vspace{0.1in} \textbf{Proof of Theorem 2:} The proof will rely on
the technical Lemma 1 which is stated next and established in
Appendix A.

\vspace{0.1in} {\em \textbf{Lemma 1:} Let $G = (V,T)$ be a tree,
and associate with each edge a bit.  Then the terminals in $V$ can
devise a (noninteractive) LC of length $|T| - 1$ bits enabling
every terminal in $V$ to recover all the edges of $T$, i.e., all
the bits associated with the edges of $T$}. \vspace{0.1in}

(i,ii) If $\mu(A, G^{(n)}) = k$, say, then $E^{(n)}$ is the
disjoint union of $k$ Steiner trees $T_1, \ldots, T_k$ (each of
which covers $A$) and the remaining edge set $R$, so that
\begin{equation}
    | E^{(n)} | ~=~ n | E^{(1)} | ~=~ \sum_{i=1}^k | T_i | ~+~ | R |, \label{eqn:Sec4-5}
\end{equation}
where $|T_i|$ denote the number of edges in $T_i$.

Apply Lemma 1 to every Steiner tree $T_i,\ i = 1, \ldots, k$, in
(\ref{eqn:Sec4-5}) to get $k$ LCs that enable every terminal in
$A$ to recover the edges of all the $T_i,\ i=1, \ldots, k$. An
additional communication of $| R |$ bits will lead to the recovery
of the leftover edges in $R$.  Thus, there exists an
$\mbox{LCO}^{(n)}(A)$ of length
\[\sum_{i=1}^k | T_i | \, - \,\, k \, + \, | R | =
  n | E^{(1)} | \,-\,\,k\ \mbox{(bits)},\]
which establishes the first assertion of (i); also, clearly,
$\mbox{LCO}_{m}^{(n)}(A) \leq n | E^{(1)} | \, - \,\,k$, thereby
proving (ii).  To establish the second assertion of (i), it
remains to extract a perfect SK from the perfect omniscience
obtained using the $\mbox{LCO}^{(n)}(A)$ above of total length $~n
| E^{(1)} | \, - \,\,\mu(A,G^{(n)})~$ (bits).  This is
accomplished exactly as in the proof of Theorem 1, whereby the
terminals in $A$ extract a perfect SK $K^{(n)}$ with
$\log{|\mathcal{K}^{(n)}|} = \mu(A, G^{(n)})$.

(iii)  Consider an $\mbox{LCO}^{(n)}(A) = (L_1, \ldots, L_m)$
achieving $\mbox{LCO}_{m}^{(n)}(A)$ with $(b_1, \ldots, b_m)$
(bits), respectively.  Fix $B \subset \mathcal{M},\ B \nsupseteq
A$, and consider $\mathcal{S}=\{\tilde{x}^n_{\mathcal{M}}:\
\tilde{x}^n_{j}=\mathbf{0}\ \mbox{for~every}~j \in B^c\}$ with
cardinality $2^{n \sum_{1 \leq i < j \leq m,\ i \in B,\ j \in B} ~
e_{ij}}$. For every $k \in B^c \cap A$ and every
$\tilde{x}^n_{\mathcal{M}} \in \mathcal{S}$, it holds that
$\tilde{x}^n_{k} = \mathbf{0}$. Consequently, by the perfect
recoverability property of an $\mbox{LCO}^{(n)}(A)$, such a
terminal $k$ must be able to discern all the sequences in
$\mathcal{S}$ using {\em only} $(L_1, \ldots, L_m)$. Note also
that for every $\tilde{x}^n_{\mathcal{M}} \in \mathcal{S}$ and
every $i \in B^c$, it follows that $L_i(\tilde{x}^n_{i}) =
\mathbf{0}$; therefore, the set of all communication messages
corresponding to $\mathcal{S}$ has cardinality at most $2^{\sum_{i
\in B} b_i}$. From the mentioned condition on perfect
recoverability at terminal $k \in B^c \cap A$ of all sequences in
$\mathcal{S}$, it must hold that $2^{\sum_{i \in B} b_i} \geq 2^{n
\sum_{1 \leq i < j \leq m,\ (i,j) \in B} ~e_{ij}}$. Since this
argument is valid for every $B \subset \mathcal{M},\ B \nsupseteq
A$, we have that $(b_1, \ldots, b_m) \in \mathcal{I}_{G^{(n)}}(A)$
and, hence, $\mbox{LCO}_{m}^{(n)}(A)$ is at least $\min_{(I_1,
\ldots,
I_m)\,\in\,\mathcal{I}_{G^{(n)}}(A)}~\sum_{i\,=\,1}^m\,I_i. \qed$

\vspace{0.1in} \textbf{Proof of Corollary 3:} The inequality in
the Corollary $3$ is immediate from (\ref{eqn:Sec3.2-7-2}) and
(\ref{eqn:Sec3.2-9}). Equality when $A = \mathcal{M}$ relies on
Lemma 2 and 3 below; Lemma 2 is a classic result of Nash-Williams
\cite{Nash61} and Tutte \cite{Tutte61} on the maximal size of
spanning tree packing of a multigraph, and Lemma 3 \cite{CsiNar04}
provides an upper bound for (standard) SK capacity

\vspace{0.1in} {\em \textbf{ Lemma 2:} \cite{Nash61},
\cite{Tutte61} For a multigraph $G = (\mathcal{M}, E)$,
\begin{equation}
\mu(\mathcal{M}, G) ~=~ \Big \lfloor \min_{\mathcal{P}} ~
    \frac{1}{|\mathcal{P}|-1} ~ \Big | \{e \in E: e \mbox{~crosses~}\mathcal{P} \}
                        \Big |
    \Big \rfloor, \nonumber
\end{equation}
where the minimum is over all partitions $\mathcal{P}$ of
$\mathcal{M}$.}

\vspace{0.1in}
{\em \textbf{Lemma 3:} \cite{CsiNar04}  For the
multigraph $G = (\mathcal{M}, E)$ associated with the PIN model
and for $A \subseteq \mathcal{M}$,
\[C(A) = |E| - \min_{(R_1, \ldots, R_m)~\in~\mathcal{R}_G(A)}\
    \sum_{i\,=\,1}^m\,R_i
\leq \]
\[\ \ \ \min_{\mathcal{P}} ~
    \frac{1}{|\mathcal{P}|-1} ~ \Big |\{e \in E: e \mbox{~crosses~}\mathcal{P} \} \Big |,
\]
where the minimum is over all partitions $\mathcal{P}$ of
$\mathcal{M}$ such that each atom of $\mathcal{P}$ intersects $A$
.}

\vspace{0.1in}
By (\ref{eqn:Sec3.2-5}) and (\ref{eqn:Sec3.2-9}),
$\mathcal{R}_{G^{(n)}}(\mathcal{M}) \supset
\mathcal{I}_{G^{(n)}}(\mathcal{M})$ with $G^{(n)}$ and
$\mathcal{M}$ in the roles of $G$ and $A$ in (\ref{eqn:Sec3.2-5}),
it is clear that
\begin{eqnarray}
\Big \lceil
 \min_{(R_1, \ldots, R_m)~\in~\mathcal{R}_{G^{(n)}}(\mathcal{M})}\
    \sum_{i\,=\,1}^m\,R_i
\Big \rceil  \nonumber \\
\ \ \ \ \leq \min_{(I_1, \ldots,
I_m)~\in~\mathcal{I}_{G^{(n)}}(\mathcal{M})}\
    \sum_{i\,=\,1}^m\,I_i, \label{eqn:Sec4-6}
\end{eqnarray}
noting that the value on the right-side above is an integer.

Then the claimed equality follows since

\vspace{0.05in} $\mu(\mathcal{M}, G^{(n)})$
\begin{eqnarray}
    &\leq&  n~|E^{(1)}| - \min_{(I_1, \ldots, I_m)~\in~\mathcal{I}_{G^{(n)}}(\mathcal{M})}\
        \sum_{i\,=\,1}^m\,I_i \nonumber \\
    &\leq& \Big \lfloor n~|E^{(1)}| - \min_{(R_1, \ldots, R_m)~\in~\mathcal{R}_{G^{(n)}}(\mathcal{M})}\
    \sum_{i\,=\,1}^m\,R_i \Big \rfloor, \  \ \  \mbox{by}\ (\ref{eqn:Sec4-6}) \nonumber \\
    &\leq& \Big \lfloor \min_{\mathcal{P}} ~
    \frac{1}{|\mathcal{P}|-1} ~ \Big |\{e \in E^{(n)}: e \mbox{~crosses~}\mathcal{P} \} \Big |
    \Big \rfloor \label{eqn:Sec4-7}
\end{eqnarray}
\begin{eqnarray}
    &=& \mu(\mathcal{M}, G^{(n)}),\ \ \mbox{by~Lemma~2}, \nonumber
\end{eqnarray}
where (\ref{eqn:Sec4-7}) is by Lemma 3.  $\sqr$

\vspace{0.1in} \textbf{Proof of Proposition 4:} By remark (ii)
after Definition 7 in section III, we have that
\begin{eqnarray}
    \frac{1}{n}\mu(A, G^{(n)})
    &=& \frac{1}{n} \max_{\mathcal{T}_G^{(n)}(A) \cap \mathbb{Z}^k }~
        \sum_{l = 1}^k  T_l \nonumber \\
    &=& \max_{\mathcal{T}_G(A) \cap \frac{1}{n}\mathbb{Z}^k }~
        \sum_{l = 1}^k  T_l. \nonumber
\end{eqnarray}
Since
\begin{equation}
    \lim_{n \rightarrow \infty} \max_{\mathcal{T}_G(A) \cap \frac{1}{n}\mathbb{Z}^k }~
        \sum_{l = 1}^k  T_l
    ~=~ \max_{\mathcal{T}_G(A)}
    \sum_{l = 1}^k  T_l ~=~  \mu_f(A,G),\nonumber \
\end{equation}
the assertion follows. $\sqr$

\vspace{0.1in} \textbf{Proof of Theorem 5:} The second inequality
of the theorem is immediate by Theorem 2 (i) and the definition of
$C(A)$.

The proof of the first inequality takes recourse to the following
result.

\vspace{0.1in}
{\em \textbf{Lemma 4:} \cite{Lovasz, Kriesell} For
a multigraph $G=(\mathcal{M}, E)$ that is Eulerian\footnote{The
number of edges incident on each vertex is even.} and $A \subseteq
\mathcal{M}$,
\[\mu(A, G) \geq
    \Big \lfloor \frac{1}{2}~\min_{C \subset \mathcal{M}:C \cap A \neq \emptyset}
            ~ \Big |\{e \in E: e \mbox{~crosses~} C, C^c \} \Big | \Big \rfloor.
\]
}

\vspace{0.1in}
Now, for every $n$, $\mathcal{R}_{G^{(n)}}(A)
\supset \mathcal{I}_{G^{(n)}}(A)$, and so
\[\min_{\mathcal{I}_{G^{(n)}}(A)}\
    \sum_{i\,=\,1}^m\,I_i \geq
  \min_{\mathcal{R}_{G^{(n)}}(A)}\
    \sum_{i\,=\,1}^m\,R_i.
\]
By Lemma 3,
\[\hspace{-1.9in} n |E^{(1)}| -  \min_{\mathcal{R}_{G^{(n)}}(A)}\
                        \sum_{i\,=\,1}^m\,R_i\]
\vspace{-0.3in}
\begin{eqnarray}
    &\leq& \min_{\mathcal{P}}
            \frac{1}{|\mathcal{P}|-1} ~ \Big |\{e \in E^{(n)}: e \mbox{~crosses~}\mathcal{P} \} \Big | \nonumber\\
    &\leq& \min_{C \subset \mathcal{M}:C \cap A \neq \emptyset}
            \Big |\{e \in E^{(n)}: e \mbox{~crosses~} C, C^c \} \Big |.
    \label{eqn:Sec4-8}
\end{eqnarray}
Restricting ourselves to $n$ even, note that $G^{(n)}$ is
Eulerian, i.e., each vertex has even degree.  Then since the term
within $\lfloor ~\rfloor$ in the right side in Lemma 4 is clearly
an integer, we have that

\vspace{0.05in} $\mu(A,G^{(n)})$
\begin{eqnarray}
    &\geq& \frac{1}{2}~\min_{\emptyset \neq C \subset \mathcal{M}:C \cap A \neq \emptyset}
            ~ \Big |\{e \in E^{(n)}: e \mbox{~crosses~} C, C^c \} \Big |  \nonumber \\
    &\geq& \frac{1}{2}  \left [ n |E^{(1)}| -  \min_{\mathcal{R}_{G^{(n)}}(A)}\
                        \sum_{i\,=\,1}^m\,R_i  \right ],\ \ \mbox{by~}(\ref{eqn:Sec4-8}) \nonumber \\
    &=& \frac{1}{2}  \left [n |E^{(1)}| -  OMN_{G^{(n)}}(A)  \right ] \nonumber \\
    &=& \frac{1}{2}  n \left [|E^{(1)}| -  OMN_G(A) \right ], \ \ \mbox{by~}(\ref{eqn:Sec3.2-7}) \nonumber \\
    &=& \frac{1}{2}  n C(A), \nonumber
\end{eqnarray}
thereby establishing the left inequality of the theorem.
$\blacksquare$

\vspace{0.1in} \textbf{Proof of Proposition 6:} We prove (i) and
(ii). The proofs of (iii) and (iv) are similar but simpler, and
are omitted.

(i) Similarly as in remark (i) following Definition 7, we note
that the right-side of (i) corresponds to a linear program with
finite optimum value, and the maximum is attained.  Let
$\tilde{G}^*$, $(T_1^*, \ldots, T_{k_1}^*)$, $(T_1^{**}, \ldots,
T_{k_2}^{**})$ attain the maximum in the right side of (i), where
$(T_1^*, \ldots, T_{k_1}^*)$ and $(T_1^{**}, \ldots,
T_{k_2}^{**})$ attain the respective maxima in $\mu_f(A,
\tilde{G}^*)$ and $\mu_f(\mathcal{M}, G \backslash \tilde{G}^*)$,
with $k_1$ (resp. $k_2$) being the number of all distinct spanning
trees in $A$ (resp. $\mathcal{M}$) of $G$.  Clearly, $(T_1^*,
\ldots, T_{k_1}^*, T_1^{**}, \ldots, T_{k_2}^{**})$ is feasible
for $\mu_f(A,G)$, noting that a Steiner tree for $A$ of $G$ is
either a spanning tree in $A$ or a spanning tree in $\mathcal{M}$.

(ii) Similarly as in the proof of (i), we let $\tilde{G}^*$
$(R_1^*, \ldots, R_{m-1}^*)$, $(R_1^{**}, \ldots, R_{m}^{**})$
attain the minimum in the right side of (ii), where $(R_1^*,
\ldots, R_{m-1}^*)$ and $(R_1^{**}, \ldots, R_{m}^{**})$ attain
the respective minima in $OMN_{\tilde{G}^*}(A)$ and $OMN_{G
\backslash \tilde{G}^*}(\mathcal{M})$.  Clearly, $(R_1^* +
R_1^{**}, \ldots, R_{m-1}^* + R_{m-1}^{**}, R_m^{**})$ is feasible
for $OMN_G(A)$, thereby proving (ii).

Similar arguments considering the corresponding integer linear
programs lead to (iii) and (iv).  $\sqr$

\vspace{0.1in} \textbf{Proof of Theorem 7:} We shall prove only
(i); the proof of (ii) is similar and is omitted.

First, we show that (\ref{eqn:Sec3.3-16}) implies
(\ref{eqn:Sec3.3-15}), i.e.,
\begin{equation}
    \lim_{n \rightarrow \infty} \frac{1}{n} \mu(A, G^{(n)}) \geq C(A) = |E| -
    OMN_G(A), \label{eqn:Sec4-9}
\end{equation}
(since the reverse inequality always hold by Theorem 5).  Let a
fractional multigraph $\tilde{G}^*=(A, \tilde{E}^*)$ achieve the
minimum in the right side of (\ref{eqn:Sec3.3-16}). Then,
\begin{eqnarray}
    \lim_{n \rightarrow \infty} \frac{1}{n} \mu(A,G^{(n)})
        &=& \mu_f(A,G),\ \ \mbox{by~(\ref{eqn:Sec3.2-13})} \nonumber \\
        &\geq& \max_{\tilde{G}}\ \mu_f(A, \tilde{G}) + \mu_f(\mathcal{M}, G \backslash
                    \tilde{G}), \nonumber \\
        & & \ \ \ \ \ \ \ \mbox{by~Proposition~6~(i)} \nonumber \\
        &\geq& \mu_f(A, \tilde{G}^*) + \mu_f(\mathcal{M}, G \backslash
                    \tilde{G}^*). \label{eqn:Sec4-10}
\end{eqnarray}
Next, because the linear program in the right side of
(\ref{eqn:Sec3.3-16}) involves a cost and linear constraints with
only integer-valued coefficients, $\tilde{G}^*=(A, \tilde{E}^*)$
can always be taken to be {\em rational}, i.e., all
$\tilde{e}^*_{ij}$s in $\tilde{E}^*$ are rational. Next, let $l$
be the least common multiple of all $\tilde{e}^*_{ij}$s so that
$\tilde{G}^{*(l)}=(A, \tilde{E}^{*(l)})$ is a multigraph with edge
set $\tilde{E}^{*(l)} = \{l~\tilde{e}^*_{ij},\ 1 \leq i < j \leq m
- 1 \}$. Then,
\begin{eqnarray}
\mu_f(A, \tilde{G}^*) &=& \frac{1}{l} \mu_f(A, \tilde{G}^{*(l)}),\ \ \mbox{by (\ref{eqn:Sec3.2-12})} \nonumber \\
               &=& \frac{1}{l} (|\tilde{E}^{*(l)}| - OMN_{\tilde{G}^{*(l)}}(A)) \nonumber \\
               &=& |\tilde{E}^*| - OMN_{\tilde{G}^*}(A),\ \ \mbox{by (\ref{eqn:Sec3.2-7})}; \label{eqn:Sec4-11}
\end{eqnarray}
the second equality is by Proposition 4 and the second assertion
of Theorem 5 noting that the vertex set of $\tilde{G}^{*(l)}$ is
$A$. By a similar argument, we have that
\begin{equation}
\mu_f(\mathcal{M}, G \backslash \tilde{G}^*)
               = |E \backslash \tilde{E}^*| - OMN_{G \backslash \tilde{G}^*}(\mathcal{M}). \label{eqn:Sec4-12}
\end{equation}

Substituting (\ref{eqn:Sec4-11}) and (\ref{eqn:Sec4-12}) in
(\ref{eqn:Sec4-10}),

$\lim_{n \rightarrow \infty} \frac{1}{n} \mu(A,G^{(n)})$
\begin{eqnarray}
    &\geq& |\tilde{E}^*| + |E \backslash \tilde{E}^*| \nonumber \\
    & &\ - ( OMN_{\tilde{G}^*}(A) + OMN_{G \backslash \tilde{G}^*}(\mathcal{M}) ) \nonumber \\
    &=& |E| - OMN_{G}(A),\ \ \mbox{by}~(\ref{eqn:Sec3.3-16}) \nonumber
\end{eqnarray}
thereby giving (\ref{eqn:Sec4-9}).

Conversely, to prove that (\ref{eqn:Sec3.3-15}) implies
(\ref{eqn:Sec3.3-16}), i.e.,
\begin{equation}
    {OMN}_G(A) \geq \min_{\tilde{G}} {OMN}_{\tilde{G}}(A) +
    {OMN}_{G \backslash \tilde{G}}(\mathcal{M})    \nonumber
\end{equation}
(since the reverse inequality always holds by Proposition 6 (ii)),
we can assume similarly as above that $\mu_f(A, G)$ is attained by
$(T_1^*, \ldots, T_k^*)$ with rational components, where $k=k(G)$
is the number of distinct Steiner trees (for $A$) of $G$ (see
passage preceding (\ref{eqn:Sec3.2-11})). Next, since $A = \{1,
\ldots, m-1\} \subset \mathcal{M}$, the collection of all distinct
Steiner trees of (for $A$) of $G$, namely $\{S_1, \ldots, S_k\}$
can be decomposed as $\mathcal{S}_1 \sqcup \mathcal{S}_2$, where
$\mathcal{S}_1$ (resp. $\mathcal{S}_2$) comprises all spanning
trees in $A$ (resp. $\mathcal{M}$). Consider the fractional
multigraph in $A$ defined by
\begin{equation}
    \tilde{\tilde{G}}^* = (A, \tilde{\tilde{E}}^*),\ \tilde{\tilde{E}}^*\ =\{\tilde{\tilde{e}}^*_{ij} = \mathop{\sum_{l: (i,j) \in S_l,}}_{~~S_l
\in \mathcal{S}_1} T_l^*, 1 \leq i < j \leq m-1 \}) \nonumber
\end{equation}
Then, it follows that
\begin{equation}
\mu_f(A,G) = \mu_f(A, \tilde{\tilde{G}}^*) + \mu_f(\mathcal{M}, G
\backslash \tilde{\tilde{G}}^*) \label{eqn:Sec4-13}
\end{equation}
since
\begin{eqnarray}
    \mu_f(A,G) &=& \sum_{l = 1}^k T_l^* \nonumber \\
               &=& \sum_{l:~S_l \in \mathcal{S}_1} T_l^* + \sum_{l:~S_l \in \mathcal{S}_2} T_l^* \nonumber \\
               &\leq& \mu_f(A, \tilde{\tilde{G}}^*)
                        + \mu_f(A, G \backslash \tilde{\tilde{G}}^*),
                        \nonumber
\end{eqnarray}
by the definition of $\mu_f$; the reverse inequality is always
true.  Finally, the right side of (\ref{eqn:Sec3.3-15}) satisfies

$OMN_{\tilde{G}^*}(A) + OMN_{G \backslash
\tilde{G}^*}(\mathcal{M})$
\begin{eqnarray}
     &\leq& OMN_{\tilde{\tilde{G}}^*}(A) + OMN_{G \backslash \tilde{\tilde{G}}^*}(\mathcal{M})
                \nonumber \\
     &=& ( |\tilde{\tilde{E}}^*| - \mu_f(A, \tilde{\tilde{G}}^*)) + \nonumber \\
     & & ( |E \backslash \tilde{\tilde{E}}^*| - \mu_f(\mathcal{M}, G \backslash \tilde{\tilde{G}}^*)), \nonumber \\
     & & \ \ \ \mbox{as~in~}(\ref{eqn:Sec4-11}),~(\ref{eqn:Sec4-12}) \nonumber \\
     &=& |E| - \mu_f(A,G),\ \ \mbox{by}~(\ref{eqn:Sec4-13}) \nonumber \\
     &=& OMN_G(A), \nonumber
\end{eqnarray}
by (\ref{eqn:Sec3.3-15}), (\ref{eqn:Sec3.2-13}) and
(\ref{eqn:Sec3.2-6}). $\sqr$

\vspace{0.1in} \textbf{Proof of Theorem 8:}  First, we prove (ii),
and then (i) by applying (ii) to $G^{(n)} = (\mathcal{M},
E^{(n)})$ and taking appropriate limits.

The proof of (ii) entails considering a modification of $G =
(\mathcal{M}, E)$ obtained by ``edge-splitting'' at the helper
vertex $m$.  Specifically, if $G$ has more than one vertex in $A$
connecting to $m$, then for any two such vertices $u, v \in A$,
let $G^{uv}=(\mathcal{M}, E^{uv})$ denote the multigraph obtained
from $G$ by splitting off the edges $(u, m)$ and $(v, m)$, i.e.,
by reducing $e_{um}$ and $e_{vm}$ each by unity and increasing
$e_{uv}$ by unity; note that $|E^{uv}| = |E| - 1$.

The following claim, whose proof is relegated to Appendix B, will
be used to establish the theorem.

{\em Claim: For a multigraph $G = (\mathcal{M}, E)$,

(a) if $m$ is connected to at most one vertex in $A$ or if there
exists $(I_1^*, \ldots, I_m^*)$ attaining $INT_G(A)$ with $I_m^* =
0$, then
\begin{equation}
    \mu(A, G) = |E| - INT_G(A);
    \label{eqn:Sec4-14}
\end{equation}

(b) if $m$ is connected to more than one vertex in $A$ and if
there exists $(I_1^*, \ldots, I_m^*)$
attaining $INT_G(A)$ with $0 < I_m^* \leq \lfloor
d_m/2 \rfloor$, then for $u \in A$ connecting to $m$ there exists
$v = v(u) \in A,\ v \neq u$, also connecting to $m$, such that
$(I_1^*, \ldots, , I_{m-1}^*, I_m^*-1)$ attains $INT_{G^{uv}}(A)$,
and so
\begin{equation}
    |E| - INT_G(A) = |E^{uv}| - INT_{G^{uv}}(A);
    \label{eqn:Sec4-15}
\end{equation}

(c) if $m$ is connected to more than one vertex in $A$, then for
$u, v \in A$ both connecting to $m$,
\begin{equation}
    \mu(A, G) \geq \mu(A, G^{uv}). \nonumber
\end{equation}
}

In order to prove (ii), we observe first that it holds if the
hypothesis of Claim (a) is met.  It remains to consider the realm
of Claim (b).  Let $(I_1^*, \ldots, I_m^*)$ be as in Claim (b).
Then we obtain $G_2 = (\mathcal{M}, E_2) = G^{u v}$ for some $u, v
\in A$ connecting to $m$, and with $(I_1^*, \ldots, I_m^*-1)$
attaining $INT_{G_2}(A)$.  If $I_m^* - 1 = 0$ or $m$ connects to
at most one vertex in $A$ in $G_2$, then by (\ref{eqn:Sec4-14})
(\ref{eqn:Sec4-15}),
\[\mu(A, G_2) ~=~ |E_2| - INT_{G_2}(A) ~=~ |E| - INT_{G}(A).\]
Else, $G_2 = (\mathcal{M}, E_2)$ is back in the realm of Claim
(b), noting that the degree of $m$ in $G_2$ is $d_m - 2$ and
$I_m^* - 1 \leq \lfloor (d_m-2)/2 \rfloor$ as $2 \leq I_m^* \leq
\lfloor d_m/2 \rfloor$.\\
Thus, we obtain a finite number of multigraphs $G_1 = G, G_2,
\ldots, G_q$, such that $G_i = (\mathcal{M}, E_i) = G_{i-1}^{u
v}~$ for some $(u, v)=(u, v)(i)$ in $A$, and satisfying
\begin{equation}
    |E_{i-1}| - INT_{G_{i-1}}(A) = |E_{i}| - INT_{G_{i}}(A),\ i = 2, \ldots, q
    \label{eqn:Sec4-16}
\end{equation}
and
\begin{equation}
    \mu(A, G_q) = |E_q| - INT_{G_q}(A).
    \label{eqn:Sec4-17}
\end{equation}
Using Claim (c) repeatedly,
\begin{eqnarray}
    \mu(A, G) &=& \mu(A, G_1) \geq \mu(A, G_q) \nonumber \\
        &=& |E_q| - INT_{G_q}(A).\ \ \mbox{by}~(\ref{eqn:Sec4-17}) \nonumber\\
        &=& |E| - INT_{G}(A)    \label{eqn:Sec4-18}
\end{eqnarray}
by the repeated use of (\ref{eqn:Sec4-16}).  Then, (ii) is
immediate from (\ref{eqn:Sec4-18}) and Corollary 3.

To establish (i), the hypothesis implies (with a slight abuse of
notation) that
\begin{equation}
    \min_{\mathcal{R}_G(A) \bigcap \{R_m \leq d_m/2\}} ~
    \sum_{i\,=\,1}^m\,R_i = OMN_G(A). \label{eqn:Sec4-19}
\end{equation}
Pick $(R_1^*, \ldots, R_m^*)$ that attains the left side with all
rational components, and let $l$ be the least common multiple of
their denominators.  Thus, for every integer $n \geq 1$, $(n l
R_1^*, \ldots, n l R_m^*)$ attains $INT_{G^{(nl)}}(A)$.  As $n l
R_m^* \leq n l \frac{d_m}{2}$, it follows from (ii) that
\begin{eqnarray}
    \mu(A, G^{(nl)}) &=&  n l |E| - INT_{G^{(nl)}}(A) \nonumber \\
                   &=& n l |E| - n l OMN_G(A),\ \ \mbox{by}~(\ref{eqn:Sec4-19}).  \nonumber
\end{eqnarray}
Upon dividing both sides by $nl$ and taking limits as $n
\rightarrow \infty$ (with $l$ fixed), we obtain (i).
$\blacksquare$

\section{Discussion}

We conclude by mentioning several unresolved questions
raised by this work.

When all the terminals in $\mathcal{M}$ see to share a perfect SK,
i.e., $A = \mathcal{M}$, we see from Theorem 5 that maximal
spanning tree packing attains perfect SK capacity; this is no
longer true, in general, when $A \subset \mathcal{M}$ (cf. the
example in section III.B).  However, the single helper model in
section III.C possesses the special feature that a Steiner tree for
$A$ is a spanning tree for either $A$ or $\mathcal{M}$.  In spite
of this, it is unresolved whether a maximal Steiner tree packing
of $A$ attains perfect SK capacity (i.e., if the second inequality
in (\ref{eqn:Sec3.2-14}) is tight) or if (\ref{eqn:Sec3.2-10})
holds with equality (whereupon the sufficient conditions of
Theorem 8 become superfluous).  We note that the optimality of
maximal spanning tree packing in (\ref{eqn:Sec3.2-10}) and
(\ref{eqn:Sec3.2-14-b}), constitutes, in effect, a reformulation
of the classic graph-theoretic results of Nash-Williams
\cite{Nash61} and Tutte \cite{Tutte61}.  A better {\em information
theoretic} understanding of (\ref{eqn:Sec3.2-10}) and
(\ref{eqn:Sec3.2-14-b}) is desirable, and might suggest
alternative interpretations of related results in combinatorial
tree packing.

Perfect SK capacity in Theorem 1 was shown to be achievable by way
of the attainment of perfect omniscience at a minimum
communication rate $OMN(A)$.  However, when $A = \mathcal{M}$,
Theorem 5 asserts that maximal spanning tree packing attains capacity; an
examination of its proof (cf. Lemma 1) shows the corresponding
rate of communication to be $(m-1)C(\mathcal{M})$ which can be
less than $OMN(\mathcal{M})$.  It remains open to characterize the
minimum rate of public communication needed to attain perfect SK
capacity.

Maximal Steiner tree packing is guaranteed by Theorem 5 to attain
a fraction of at least half of the capacity $C(A)$.  What is the
best feasible value of this fraction?

Lastly, the design of efficient algorithm for perfect SK
generation is largely unexplored.
\\

\renewcommand{\theequation}{A-\arabic{equation}}
\setcounter{equation}{0}  
\section*{Appendix A: Proof of Lemma 1}

We prove a slightly stronger result that there exists an LC whose
null space comprises only the all-zero and the all-one strings
(corresponding to the edges in $T$ being labelled all zero or all
one) which clearly enables every terminal in $V$ to recover all
the edges of $T$.  We prove the claim by induction. When $|T| =
2$, say, with $T = \{e_1 = (v_1, v_2), e_2=(v_2, v_3)\},$ then
$e_1 + e_2 \mbox{~mod~}2$ constitutes an LC whose null space is
$\{(00), (11)\}$. Next, suppose the claim is true for all trees
with $k-1$ edges, $k \geq 3$.  Given a tree with $k$ edges, pick
an end vertex $v_{k+1}$ of the tree (a vertex with degree one),
and let $v_k$ be the sole vertex connecting to $v_{k+1}$. Then
$G=(V, T' \bigcup \{(v_{k}, v_{k+1})\})$, and $G'=(V \backslash
\{v_{k+1}\}, T')$ is a subtree of $G$.  By the induction
hypothesis, there exists an LC for $G'$, say, $F(T')$ of length
$k-2$ (bits) and whose null space is $\{{\bf 0}^{k-1}, {\bf
1}^{k-1}\}$. Let $v_{k-1}$ be another vertex connecting to $v_k$
and let $e_{k-1} = (v_{k-1}, v_k)$ and $e_{k} = (v_{k}, v_{k+1})$.
Then, consider $\{F(T'), e_{k-1}+e_k\}$ as an LC of $G$ of length
$k-1$. It is now clear that the null space of this LC is $\{{\bf
0}^{k}, {\bf 1}^{k}\}$. $\sqr$

\renewcommand{\theequation}{B-\arabic{equation}}
\setcounter{equation}{0}  
\section*{Appendix B: Proof of Claim in (the proof of) Theorem 8}

(a) Let $G_A=(A, E_A)$ denote a subgraph of $G$ in $A$, where $E_A
\subset E$ consists only of those edges in $E$ whose both end
vertices lie in $A$.  Clearly,
\begin{eqnarray}
|E| -  INT_G(A) &\geq& \mu(A, G) ~\geq~ \mu(A, G_A) \nonumber \\
    &=& |E_A| - INT_{G_A}(A),\ \nonumber \\
    & &\ \     \mbox{by~Corollary~}3\mbox{~with~}\mathcal{M}=A \nonumber \\
    &=& |E| - \left ( d_m + INT_{G_A}(A) \right ). \nonumber
\end{eqnarray}
Thus, it suffices to show that
\begin{equation}
    d_m + INT_{G_A}(A) \leq INT_G(A).   \label{eqn:SecAppB1}
\end{equation}
Consider first the case where $(I_1^*, \ldots, I_{m-1}^*, 0)$
attains $INT_{G}(A)$.  Without loss of generality, let $\{1,
\ldots, a\},\ a \leq m-1$, be the set of vertices in $A$
connecting to $m$.  For any $v \in \{1, \ldots, a\}$, since $\{v,
m\} \nsupseteq A$, we have that $I_v^* + I_m^* = I_v^* \geq
e_{vm}$ (see (\ref{eqn:Sec3.2-9})).  Consequently, since $d_m =
\sum_{u~=~1}^a e_{um}$, we see that $(I_1^* - e_{1m}, \ldots,
I_a^* - e_{am}, I_{a+1}^*, \ldots, I_{m-1}^*)$, with components
summing to $INT_G(A) - d_m$ is feasible for $INT_{G_A}(A)$. Thus,
$INT_{G}(A) - d_m \geq INT_{G_A}(A)$, establishing
(\ref{eqn:SecAppB1}). A nearly identical argument would show that
(\ref{eqn:SecAppB1}) holds too for the case when at most vertex 1
is connected to $m$, and is omitted.

(b)  Consider any $G^{uv} =  (\mathcal{M}, E^{uv})$ as in the
second paragraph of the proof of Theorem 8, and let $(I_1^{**},
\ldots, I_m^{**})$ attain $INT_{G^{uv}}(A)$.  Then, $(I_1^{**},
\ldots, I_{m-1}^{**}, I_m^{**}+1)$ is feasible for $INT_G(A)$, so
that
\begin{equation}
    INT_G(A) \leq INT_{G^{uv}}(A) + 1.  \label{eqn:SecAppB2}
\end{equation}
Without loss of generality, let $\{1, \ldots, a\}$ be as in the
proof of Claim (a).  To prove Claim (b), it suffices to show for
$u = 1$ that there exists $v \in \{2, \ldots, a\}$ such that
$(I_1^*, \ldots, I_{m-1}^*, I_m^*-1)$ is feasible for
$INT_{G^{1v}}(A)$ if $0 < I_m^* \leq \lfloor \frac{d_m}{2}
\rfloor$.  This would mean that
\begin{equation}
    INT_G(A) - 1 \geq INT_{G^{1v}}(A).  \label{eqn:SecAppB3}
\end{equation}
which, together with the observation that $|E|-1 = |E^{1v}|$,
establishes Claim (b).  To this end, referring to
(\ref{eqn:Sec3.2-9}), for $B \subseteq \mathcal{M}$, set
\begin{equation}
e_G(B) \triangleq \sum_{1 \leq i < j \leq m,\ i\in B,\ j \in B}
e_{ij}, \ \ \ e_G(\emptyset) \triangleq 0,  \label{eqn:SecAppB4}
\end{equation}
and let
\begin{equation}
\mathcal{B} =
\left \{ \begin{array}{ll}
   B,\ \emptyset \neq B \subset \mathcal{M},\ B \nsupseteq A,\ \\
   \sum_{i \in B} I_i^* = e_G(B)
   \end{array} \right \}.
    \label{eqn:SecAppB5}
\end{equation}
We make the following

{\em Claim (d): For $u=1$, there exists $v \in \{2, \ldots, a\}$
connecting to $m$ with the properties that

a) for $B \in \mathcal{B}$ such that $1 \notin B,\ m \in B$, it
holds that $v \in B$;

b) for $B \in \mathcal{B}$ such that $1 \in B,\ m \notin B$, it
holds that $v \notin B$.}

Then, with the choice of $v$ as in the Claim (d), a simple check
of all the possibilities for $B$ (in $\mathcal{B}$ or in
$\mathcal{B}^c$) that are feasible in (\ref{eqn:Sec3.2-9}), shows
that $(I_1^*, \ldots, I_{m-1}^*, I_m^*-1)$ is feasible for
$INT_{G^{1v}}(A)$, thereby establishing (\ref{eqn:SecAppB3}) (and
hence Claim (b)).

It only remains to establish Claim (d).  We first state the
following facts with accompanying proofs.

{\em Fact 1: For $B_1, B_2 \subset \mathcal{M},\ e_G(B_1) +
e_G(B_2) \leq e_G(B_1 \cup B_2) + e_G(B_1 \cap B_2)$.}  This holds
by observing that $e_G(B_1 \cup B_2) + e_G(B_1 \cap B_2) -
e_G(B_1) - e_G(B_2) = \sum_{1 \leq i < j \leq m,\ i \in B_1
\backslash B_2,\ j \in B_2 \backslash B_1 \mbox{~or~} i \in B_2
\backslash B_1,\ j \in B_1 \backslash B_2} e_{ij} \geq 0.$

{\em Fact 2: For $B_1, B_2 \in \mathcal{B}$ with $B_1 \cup B_2
\nsupseteq A$, it holds that $B_1 \cup B_2$ and $B_1 \cap B_2$ are
both in $\mathcal{B}$.}  To see this, note first that
\begin{eqnarray}
    \sum_{i \in B_1 \cup B_2} I_i^*
        &=& \sum_{i \in B_1} I_i^* + \sum_{i \in B_2} I_i^*
                                  - \sum_{i \in B_1 \cap B_2} I_i^* \nonumber\\
        &=& e_G(B_1) + e_G(B_2)       - \sum_{i \in B_1 \cap B_2} I_i^* \nonumber\\
        &\leq& e_G(B_1) + e_G(B_2) - e_G(B_1 \cap B_2) \nonumber \\
        &\leq& e_G(B_1 \cup B_2),\ \ \mbox{by~Fact~}1. \nonumber
\end{eqnarray}
Also, $\sum_{B_1 \cup B_2} I_i^* \geq e_G(B_1 \cup B_2)$, since
$B_ 1 \cup B_2 \nsupseteq A$ is feasible in (\ref{eqn:Sec3.2-9}).
The fact follows.

{\em Fact 3: For $B \subseteq \mathcal{M}$, let $D_m(B)$ denote
the total number of edges connecting $m$ to all the vertices in $B
\cap A$.  Then, for $B \in \mathcal{B}$, if $m \in B$ then $D_m(B)
\geq I_m^*$, and if $m \notin B$ then $D_m(B) \leq I_m^*$.}  To
see this, consider first the case $m \in B \in \mathcal{B}$.  As
$\{m\} \notin \mathcal{B}$ (since $I_m^* >0$), we have $B \cap A
\neq \emptyset$.  Since $B \in \mathcal{B}$, $\sum_{i \in B} I_i^*
= e_G(B \cap A) + D_m(B)$.  Also, since $B \cap A \neq \emptyset$
is feasible in (\ref{eqn:Sec3.2-9}), $\sum_{i \in B \cap A} I_i^*
\geq e_G(B \cap A)$. Subtracting the latter from the former gives
$I_m^* \leq D_m(B)$.  The second assertion of the fact is proved
similarly.

{\em Fact 4: The intersection of all $Bs$ in $\mathcal{B}$
satisfying $1 \notin B,\ m \in B$, when nonempty, is also in
$\mathcal{B}$.  The union of all $Bs$ in $\mathcal{B}$ satisfying
$1 \in B,\ m \notin B$, when nonempty, is also in $\mathcal{B}$.}

The first assertion in Fact 4 is obtained by observing that the
union of all $Bs$ in $\mathcal{B}$ with $1 \notin B,\ m \in B$,
does not contain $A$, and by a repeated use of Fact 2. The second
assertion would follow similarly by Fact 2 if the union of all
$Bs$ in $\mathcal{B}$ with $1 \in B,\ m \notin B$, is strictly
contained in $A$.  Suppose not; then this union is exactly $A$.
The ensuing contradiction can be seen, for instance, with $B_1,
B_2$ as above with $B_1 \cup B_2 = A$.  Then
\begin{eqnarray}
    d_m &=& D_m(A) =  D_m(B_1 \cup B_2) \nonumber \\
        &=& D_m((B_1 \backslash B_2) \cup (B_1 \cap B_2) \cup (B_2
            \backslash B_1)) \nonumber \\
        &=& D_m(B_1 \backslash B_2) + D_m(B_1 \cap B_2) + D_m(B_2
            \backslash B_1) \nonumber \\
        &=& D_m(B_1) + D_m(B_2) - D_m(B_1 \cap B_2) \nonumber \\
        &\leq& I_m^* + I_m^* - 1,\ \mbox{by~Fact}~3\mbox{~and~} 1
            \in B_1 \cap B_2 \nonumber \\
        &\leq& 2 \lfloor \frac{d_m}{2} \rfloor - 1, \nonumber \\
        &  & \mbox{by~the~assumption~}I_m^* \leq \lfloor \frac{d_m}{2} \rfloor
            \nonumber \\
        &<& d_m, \nonumber
\end{eqnarray}
a contradiction.

Finally, to prove Claim (d), let $B'$ (resp. $B''$) represent the
intersection (resp. union), when nonempty, in Fact 4.  It suffices
now to show that there exists $v \in B' \cap A$ (when $B' \neq
\emptyset$) such that $v \notin B''$ and $v$ connects to $m$; this
follows from
\begin{eqnarray}
    D_m(B' \backslash B'') &=& D_m(B') - D_m(B' \cap B'') \nonumber \\
        &=& D_m(B') - (D_m(B'') - D_m(B'' \backslash B') ) \nonumber \\
        &\geq& I_m^* - (I_m^* - 1),\nonumber \\
        & & \mbox{~by~Fact}~3\mbox{~and~}1 \in B'' \backslash B' \nonumber \\
        &=& 1. \nonumber
\end{eqnarray}
Then, any $B$ as in Claim (d)(a) must contain $B'$ and hence the
$v$ above.  On the other hand, any $B$ as in Claim (d)(b) must be
contained in $B''$ and so cannot contain the $v$ above.  The cases
$B' = \emptyset$ or $B'' = \emptyset$ are handled trivially.

(c) Let $G^{uv} = (\mathcal{M}, E^{uv})$ and suppose that $T_1
\sqcup \ldots \sqcup T_k \subseteq E^{uv}$ attain $\mu(A,
G^{uv})$. If $E^{uv} \backslash \{\sqcup_{i=1}^k T_i\}$ contains
at least one edge connecting $(u,v)$, then $\{T_1, \ldots, T_k\}$
is also a Steiner tree packing of $G=(\mathcal{M}, E)$, so that
$\mu(A, G) \geq \mu(A, G^{uv})$. Else, let $T_1$, say, be the
Steiner tree that contains an edge connecting $u, v$ that emerged
by splitting off $(u, m)$ and $(v, m)$ of $G=(\mathcal{M}, E)$.
Then, $\{T_1 \backslash \{(u,v)\}\} \cup \{(u, m), (v, m)\}$ is
A-connected and hence contains a Steiner tree $T'_1$ for $A$ in
$G=(\mathcal{M}, E)$ that corresponds to $T_1$; clearly, again
$\mu(A, G) \geq \mu(A, G^{uv}).$ $\sqr$

\section*{Acknowledgement}
The authors thank Chunxuan Ye, Alexander Barg and Alex Reznik for very helpful
discussions.


\end{document}